\documentclass{emulateapj}

\newcommand{\ldl}{$\lambda/{\Delta}{\lambda}$}

\newcommand{\teff}{T$_{\rm eff}$}

\newcommand{\meth}{CH$_4$}
\newcommand{\wat}{H$_2$O}

\newcommand{\ms}{m~s$^{-1}$}
\newcommand{\kms}{km~s$^{-1}$}

\newcommand{\name}{WISE~J104915.57$-$531906.1}
\newcommand{\namesh}{Luhman~16}

\slugcomment{Submitted to ApJ 17 December 2013; { Accepted 10 February 2014}}

\shorttitle{Spectroscopic Monitoring of Luhman~16AB }
\shortauthors{Burgasser et al.}

\begin{document}

\title{A Monitoring Campaign for Luhman~16AB. I. Detection of Resolved Near-Infrared Spectroscopic Variability}

\author{
Adam J.\ Burgasser\altaffilmark{1},
Micha\"{e}l Gillon\altaffilmark{2},
Jacqueline K.\ Faherty\altaffilmark{3,4},
Jacqueline Radigan\altaffilmark{5},
Amaury H.\ M.\ J.\ Triaud\altaffilmark{6,7},
Peter Plavchan\altaffilmark{8},
Rachel Street\altaffilmark{9},
E.\ Jehin\altaffilmark{2}, 
L.\ Delrez\altaffilmark{2},
\& C.\ Opitom\altaffilmark{2}
}

\altaffiltext{1}{Center for Astrophysics and Space Science, University of California San Diego, La Jolla, CA 92093, USA; aburgasser@ucsd.edu}
\altaffiltext{2}{Institute of Astrophysics and G{\'{e}}ophysique, Universit{\'{e}} of Li{\`{e}}ge, all{\'{e}}e du 6 Ao\^{u}t, 17, B-4000 Li{\`{e}}ge, Belgium}
\altaffiltext{3}{Department of Terrestrial Magnetism, Carnegie Institution of Washington, 5241 Broad Branch Road NW, Washington, DC 20015, USA}
\altaffiltext{4}{Hubble Fellow}
\altaffiltext{5}{Space Telescope Science Institute, 3700 San Martin Drive, Baltimore, MD 21218, USA}
\altaffiltext{6}{Massachusetts Institute of Technology, Kavli Institute for Astrophysics and Space Research, 77 Massachusetts Avenue, Cambridge, MA 02139, USA}
\altaffiltext{7}{Fellow of the Swiss National Science Foundation}
\altaffiltext{8}{NASA Exoplanet Science Institute, California Institute of Technology, M/C 100-22, 770 South Wilson Avenue, Pasadena, CA 91125, USA}
\altaffiltext{9}{LCOGT, 6740 Cortona Drive, Suite 102, Goleta, CA 93117, USA}

\begin{abstract}
We report resolved near-infrared spectroscopic monitoring of the nearby L dwarf/T dwarf binary WISE~J104915.57$-$531906.1AB (Luhman~16AB), as part of a broader campaign to characterize the spectral energy distribution and temporal variability of this system.
A continuous 45-minute sequence of low-resolution IRTF/SpeX data spanning  0.8--2.4~$\micron$ were obtained, concurrent with combined-light optical photometry with ESO/TRAPPIST.  
Our spectral observations confirm the flux reversal of this binary, and we detect a wavelength-dependent decline in the relative spectral fluxes of the two components coincident with a decline in the combined-light optical brightness of the system over the course of the observation.
These data { are successfully modeled as a combination of achromatic (brightness) and chromatic (color) variability in the T0.5 Luhman~16B, consistent with variations in overall cloud opacity; and no significant variability in L7.5 Luhman~16A, consistent with recent resolved photometric monitoring.} We estimate a peak-to-peak amplitude of 13.5\% at 1.25~$\micron$ over the full lightcurve.  
Using a simple { two-spot} brightness temperature model for Luhman~16B, we infer { an average cold} covering fraction of { $\approx$30--55\%, varying by 15--30\% over a rotation period assuming a $\approx$200--400~K difference between hot and cold regions.
We interpret these variations as changes in the covering fraction of a high cloud deck and corresponding ``holes''  which expose deeper, hotter cloud layers, although other physical interpretations are possible.
A Rhines scale interpretation for the size of the variable features explains an apparent correlation between period and amplitude for Luhman~16B and the variable T dwarfs SIMP 0136+0933 and 2MASS~J2139+0220, and predicts relatively fast winds (1--3~{\kms}) for Luhman~16B 
consistent with lightcurve evolution on an advective time scale (1--3 rotation periods).
The strong variability observed in this flux reversal brown dwarf pair supports the model of a patchy disruption of the mineral cloud layer as a universal feature of the L dwarf/T dwarf transition.}   
\end{abstract}


\section{Introduction}

The driving mechanism for the transition between the L dwarf and T dwarf spectral classes has emerged as one of the outstanding problems in brown dwarf astrophysics.
Spectroscopically, this transition is defined by the { appearance} of {\meth} absorption features at near-infrared wavelengths \citep{1999ApJ...519..802K,2006ApJ...637.1067B}, accompanied by a substantial reduction of condensate cloud opacity \citep{1996Sci...272.1919M,2000ApJ...531..438B,2001ApJ...556..357A}. Both effects drive near-infrared spectral energy distributions (SEDs) to transition from red ($J-K \approx 1.5-2.5$) to blue ($J-K \approx 0-0.5$; \citealt{2000ApJ...536L..35L,2002ApJ...568..335M,2006ApJ...637.1067B}), with strengthening molecular gas bands delineating the spectral subclasses.
What is remarkable about the L dwarf/T dwarf transition is that it appears to take place 
over a relatively narrow range of effective temperatures ({\teff}s) 
and luminosities, based on 
absolute magnitude trends (e.g., \citealt{2002AJ....124.1170D,2004AJ....127.2948V}),
broad-band SED measurements (e.g., \citealt{2004AJ....127.3516G}) and
spectral model fits (e.g., \citealt{2008ApJ...678.1372C,2009ApJ...702..154S}). 
The L/T transition also exhibits
an apparent excess of binaries \citep{2007ApJ...659..655B},
gaps in color distributions \citep{2012ApJS..201...19D} 
and a decline in number densities as a function of spectral type
\citep{2008ApJ...676.1281M}, trends that suggest the transition is rapid in time as well as temperature. 

The important role of photospheric cloud evolution for this transition is seen in the observation that early-type T dwarfs with minimal cloud opacity are often significantly brighter at 1~$\micron$ than their hotter, cloudier L dwarf counterparts.
This is true in both color-magnitude
diagrams of local populations \citep{2003AJ....126..975T,2012ApJ...752...56F,2012ApJS..201...19D}
and among components of ``flux-reversal'' binaries that straddle the L/T transition
\citep{2006ApJS..166..585B,2006ApJ...647.1393L,2008ApJ...685.1183L}.
The 1~$\micron$ region is a minimum of molecular gas opacity---the local pseudocontinuum---so condensate grain scattering can dominate the overall opacity at these wavelengths \citep{2001ApJ...556..872A}.
The 1~$\micron$ brightening has thus been interpreted as a depletion of 
photospheric condensate clouds over a narrow range of {\teff} and/or time.
The geometry of the depletion has been modeled as both global changes in photospheric chemistry (e.g., \citealt{2004AJ....127.3553K,2005ApJ...621.1033T,2006ApJ...640.1063B,2008ApJ...689.1327S}) and hole formation that allows light to emerge from hotter regions \citep{2001ApJ...556..872A,2002ApJ...571L.151B,2010ApJ...723L.117M}.  
The latter hypothesis predicts an enhancement of rotationally-modulated photometric variability at the L/T transition, particularly in the 1~$\micron$ region, { depending on the sizes and distribution of the cloud gaps}. 

Recent brown dwarf monitoring observations support this prediction, as the two most prominent variables identified to date, SIMP~J013656.5+093347 (\citealt{2006ApJ...651L..57A,2009ApJ...701.1534A}; hereafter SIMP~J0136+0933) and 2MASS J21392676+0220226 (\citealt{2008AJ....136.1290R,2012ApJ...750..105R}; hereafter 2MASS~J2139+0220) are both early-type T dwarfs. Their variability can be reproduced with spot models assuming regions with thick and thin clouds at different temperatures
{ assumed to probe different layers in the atmosphere}   \citep{2012ApJ...750..105R,2013ApJ...768..121A}.  The spectral character of the observed variability is nevertheless complex.  Rather than variability being limited to pseudocontinuum regions where gas opacity is a minimum, broad-band chromatic and achromatic variations are seen across the infrared \citep{2009ApJ...701.1534A,2012ApJ...750..105R}. The light curve shapes themselves are also seen to change over several rotation periods, suggesting dynamic evolution of features at rates considerably faster than the Solar gas giants \citep{2013ApJ...776...85S}.  Finally, variability measurements over widely-separated spectral regions have recently revealed evidence of pressure-dependent phase variations, indicating vertical structure in 
{ the features driving the variability} \citep{2012ApJ...760L..31B,2013ApJ...778L..10B}.  The considerable level of detail on brown dwarf cloud structure and atmospheric dynamics garnered from these monitoring studies is of relevance to exoplanet atmospheres, where clouds are { now seen} as a key opacity source \citep{2011ApJ...733...65B,2011ApJ...737...34M,2012ApJ...754..135M,2013MNRAS.432.2917P,2013A&A...559A..33C}.

The recently-discovered, nearby binary brown dwarf system {\name}AB (hereafter {\namesh}AB; \citealt{2013ApJ...767L...1L}) has emerged as a potential benchmark for studying the L/T transition.
With spectral types of L7.5 and T0.5 \citep{2013ApJ...770..124K,2013ApJ...772..129B}, its components straddle the transition. Its T dwarf secondary is brighter than the primary in the 0.95--1.3~$\micron$ range, making it a flux-reversal system \citep{2013ApJ...772..129B}.  {\namesh}AB is also a significant variable. Combined-light red optical photometry by \citet{2013A&A...555L...5G} revealed peak-to-peak variability of $\sim$10\% with a period of 4.87$\pm$0.01~hr, with large changes in the light curve structure over daily timescales.  The variability was attributed primarily to the T dwarf component. Resolved photometry by \citet{2013ApJ...778L..10B} extended the observed variability into the near-infrared, confirmed {\namesh}B as the dominant variable, and revealed pressure-dependent phase variations.  As such, this system embodies nearly all of the remarkable characteristics of the L/T transition---multiplicity, variability, and flux reversal---while residing only 2.020$\pm$0.019~pc from the Sun \citep{2013arXiv1312.1303B}.

In April 2013, our consortium organized a week-long monitoring campaign of {\namesh}AB using telescopes in Chile, Australia and Hawaii, with the aim of characterizing its variability panchromatically (radio, optical and infrared) and spectroscopically, while simultaneously obtaining kinematic data (radial and rotational velocities) to constrain its orbit and viewing geometry.  This article reports low-resolution near-infrared spectroscopic monitoring observations obtained over 45 minutes with the SpeX spectrometer \citep{2003PASP..115..362R} on the 3.0m NASA Infrared Telescope Facility (IRTF), coincident with combined-light optical photometry obtained with the TRAnsiting Planets and PlanetesImals Small Telescope (TRAPPIST; \citealt{2011Msngr.145....2J}).
In Section~2 we describe our observation and data reduction procedures, including period analysis of the TRAPPIST lightcurve around this epoch.
In Section~3 we describe our spectral extraction and variability analysis of the SpeX data, and create an empirical model to replicate both the SpeX and TRAPPIST observations.  
In Section~4 we discuss our results, examining the nature of {\namesh}B's inferred variability in the context of a simple { two-spot} brightness temperature model, and compare this source to other significantly variable L/T transition objects.  
We summarize our results in Section~5.

\section{Observations}

\subsection{IRTF/SpeX Spectroscopy}

{\namesh}AB was observed with IRTF/SpeX on 26 April 2013 (UT) in clear and dry conditions with variable seeing.
We used the 0$\farcs$5 slit and prism-dispersed mode to obtain {\ldl} $\approx$ 120 spectra covering 0.7--2.5~$\micron$.
The source was monitored for just over an hour, between UT times 06:05 and 07:10, while seeing ranged from 0$\farcs$8 to $>$2$\arcsec$.
The slit was aligned along the binary axis at a position angle of 313$\degr$ (east of north) to obtain simultaneous spectroscopy; note that this differed significantly from the parallactic angle of {\namesh}AB, which varied from 12$\degr$ through 0$\degr$ and back to 4$\degr$ during the sequence.  
We obtained 70 exposures of 30~s each in an ABBA dither pattern.
The source never exceeded an elevation of 17$\degr$ above the horizon, and the airmass ranged from a maximum of 3.565 to a minimum of 3.420 at UT 06:51, then back up to 3.438 at the end of the sequence.  
The choice of a narrow slit was driven by guiding considerations, which was done on spill-over light from the primary using the $H$+$K$ notch filter.\footnote{This filter spans 1.47--2.4~$\micron$ with a transmission notch at 1.8~$\micron$ to block out the telluric H$_2$O absorption band; see \url{http://www.ifa.hawaii.edu/$\sim$tokunaga/filterSpecs.html}.}
For calibration, we observed the A0 V star HD~87760 ($V$ = 7.89) prior to the monitoring run at an airmass of 3.277 and with the slit aligned to the same (non-parallactic) position angle. Internal flat field and Ar arc lamp exposures were obtained 
for pixel response and wavelength calibration.

We performed an initial extraction of the combined-light spectrum as described in \citet{2013ApJ...772..129B}, using SpeXtool \citep{2003PASP..115..389V,2004PASP..116..362C} with standard settings but a wide spatial aperture that encompassed both sources.
This coarse extraction provided the wavelength calibration, telluric correction and relative flux calibration files necessary for subsequent component extractions. The combined light spectrum is slightly redder ($\Delta(J-K) = 0.1$) than that reported in \citet{2013ApJ...772..129B}, which may reflect slit losses from differential color refraction or intrinsic variability.  Detailed extraction of the component spectra are described below.

\subsection{TRAPPIST Imaging}

Throughout the overall campaign, {\namesh}AB was monitored with TRAPPIST, a 0.6~m robotic telescope located at La Silla Observatory in Chile. The telescope is equipped with a thermoelectrically-cooled 2K$\times$2K CCD camera with a 0$\farcs$65 pixel scale and a 22$\arcmin\times$22$\arcmin$ field of view. The camera images through a broad-band I + z filter with $>$90\% transmission from 0.75---1.1$\micron$, the long-wavelength cutoff set by the quantum efficiency of the CCD detector.  {\namesh}AB was observed for roughly 7.5~hr on 26 April 2013 (UT).
Data were reduced as described in \citet{2013A&A...555L...5G}.
After a standard pre-reduction (bias, dark, flatfield correction),  aperture photometry was performed using IRAF/DAOPHOT2 \citep{1987PASP...99..191S} with an aperture radius of 8 pixels (5$\farcs$2) that encompassed both sources.  Differential photometry was determined by comparison to a grid of  non-varying background stars, and the resulting light curve normalized. 

\begin{figure}[h]
\center
\epsscale{1.0}
\plotone{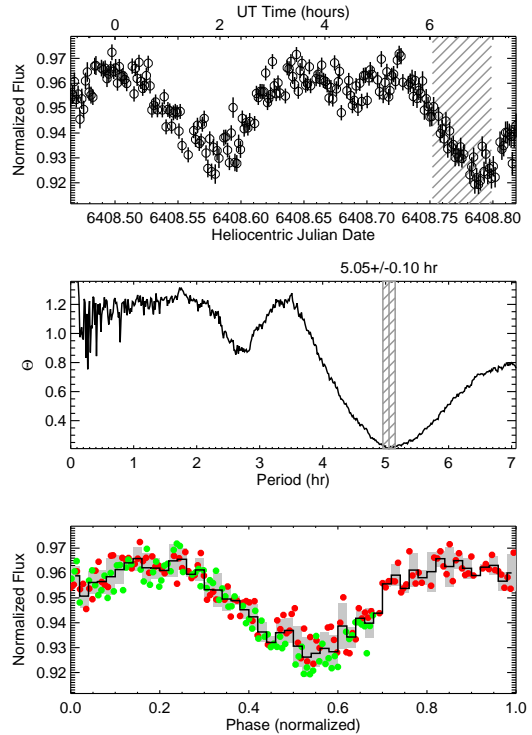}
\caption{(Top) TRAPPIST light-curve for the combined system of {\namesh}AB on 26 April 2013 (UT).  Flux values have been normalized to a global maximum.  The cross-hatched region indicates the period during which SpeX data were obtained.
(Middle) Phase-dispersion deviation statistic $\Theta \equiv \chi^2(p)/\chi^2_{sh}$ for TRAPPIST data phase folded over period $p$.  A broad minimum is found at 5.05$\pm$0.10~hr, the uncertainty determined by randomly sampling the measurement uncertainties.
(Bottom) Phase-folded light curve at 5.05~hr, with the two cycles observed indicated as red and green points.  The black histogram traces the mean light curve sampled at 50 phase points, while the grey bars indicate the 1$\sigma$ scatter of the datapoints at that phase.
}
\label{fig:trappist}
\end{figure}

That light curve is shown in Figure~\ref{fig:trappist}. As in the original
detection, {\namesh}AB exhibits significant variability over the observing period with a peak-to-peak amplitude of 5\% in a roughly sinusoidal pattern.  To determine the variability period, we used the phase dispersion minimization technique \citep{1978ApJ...224..953S}, cycling through 500 periods linearly spaced between 7.0~min (5 times the minimum sampling) and 7.1~hr (80\% of the full observational period). For each period, we phase-folded the light-curve, computed a mean curve sampled at 50 phase points across the period, then computed the $\chi^2$ deviation of the phased data from the mean curve.  We performed the same analysis with the data randomly shuffled 100 times to compute a baseline deviation ($\chi^2_{sh}$).  Figure~\ref{fig:trappist} shows that ratio of these deviations as a function of period, $\Theta = \chi^2(p)/\chi^2_{sh}$, which exhibits a broad minimum\footnote{The uncertainty of the best-fit period was determined by computing the uncertainty in $\Theta$ at each period by varying the photometric data about the measurement uncertainties 100 times. The range of periods for which $\Theta$ is within 1$\sigma$ of the minimum value set the period uncertainty.}
at 5.05$\pm$0.10~hr, longer than but statistically consistent (1.8$\sigma$) with the period measurement of \citet{2013A&A...555L...5G}.  The phased lightcurve repeats over the 1.5 periods observed, consistent with rotationally-modulated surface structure.

\section{Spectral Variability Analysis}

\subsection{Extraction of Component Spectra}

Given the seeing conditions during our SpeX observations, the component spectra are blended to varying degrees at all wavelengths.  In order to robustly separate the spectra, we directly modeled the individual data frames. We first pairwise-subtracted the raw frames, dividing each by the median-combined flat-field frame generated by SpeXtool, and then excised 48-pixel (7$\farcs$2) regions from each image along the spatial direction that encompassed both component spectral traces.  For these subimages, we performed a column-by-column fit of the spatial profiles with a six-component gaussian model: a central gaussian and two satellites for each component, with the satellites constrained to have the same separations and relative peaks for both components. All gaussians were forced to have the same widths to reduce parameter degeneracies, and each three-gaussian component profile was allowed to vary independently in amplitude and position.  Including a constant background value, this 10-parameter model was initialized by fitting to an integrated profile (summing all columns corresponding to wavelengths 1.0--1.3~$\micron$, 1.55--1.75~$\micron$ and 2.05--2.3~$\micron$) and then fitting each column individually starting from the integrated profile parameters. 
The fits were converged using an implementation of the Nelder-Mead simplex algorithm (AMOEBA; \citealt{nelder65,1986nras.book.....P}) to minimize the reduced chi-square statistic,
\begin{equation}
\chi^2_r = \frac{1}{N_{UM}-10}\sum_iW_i\frac{(P_i-M_i)^2}{\sigma_i^2},
\end{equation}
between the spatial profile $P$ and model $M$, scaled by the image variance $\sigma^2$.  A masking vector $W$ was determined by repeating the fit three times and excluding highly deviant pixels ($>$3$\sigma$), resulting in $N_{UM}$ unmasked pixels in a given column. Component fluxes at each image column were integrated directly from the final profile model, and flux uncertainties ($\sigma_{\lambda}$) were determined as
\begin{equation}
\sigma^2_{\lambda} = N_{eff}\langle\sigma^2\rangle\chi^2_r = \frac{\sum_i{M_i}}{\rm max(\{M\})} \langle\sigma^2\rangle\chi^2_r
\end{equation}
where $N_{eff}$ is the effective number of pixels used to determine the flux based on the model, and $\langle\sigma^2\rangle$ is the standard deviation between model and image counts for unmasked pixels in the spatial profile. This combination is multiplied by $\chi^2_r$ to account for systematic deviations in the profile model.

\begin{figure*}[t]
\center
\epsscale{1.0}
\plotone{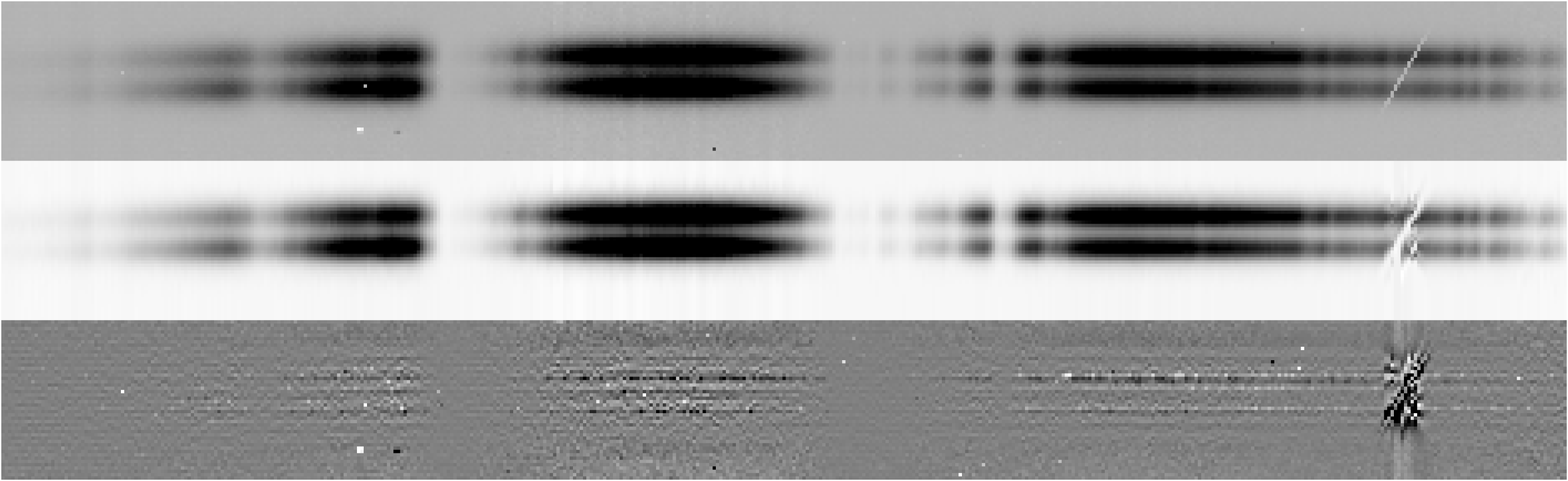}
\caption{Illustration of the forward-modeling extraction of one spectral data frame.  The top panel displays the pair-wise subtracted and flat-fielded data frame at UT time 06:02:36.
The middle panel displays the model data frame generated from the profile-fitting described in the text.  The bottom panel shows the residual image, with contrast scaled up by a factor of 10 compared to the other two frames.
}
\label{fig:model_image}
\end{figure*}

\begin{figure}[h]
\center
\plottwo{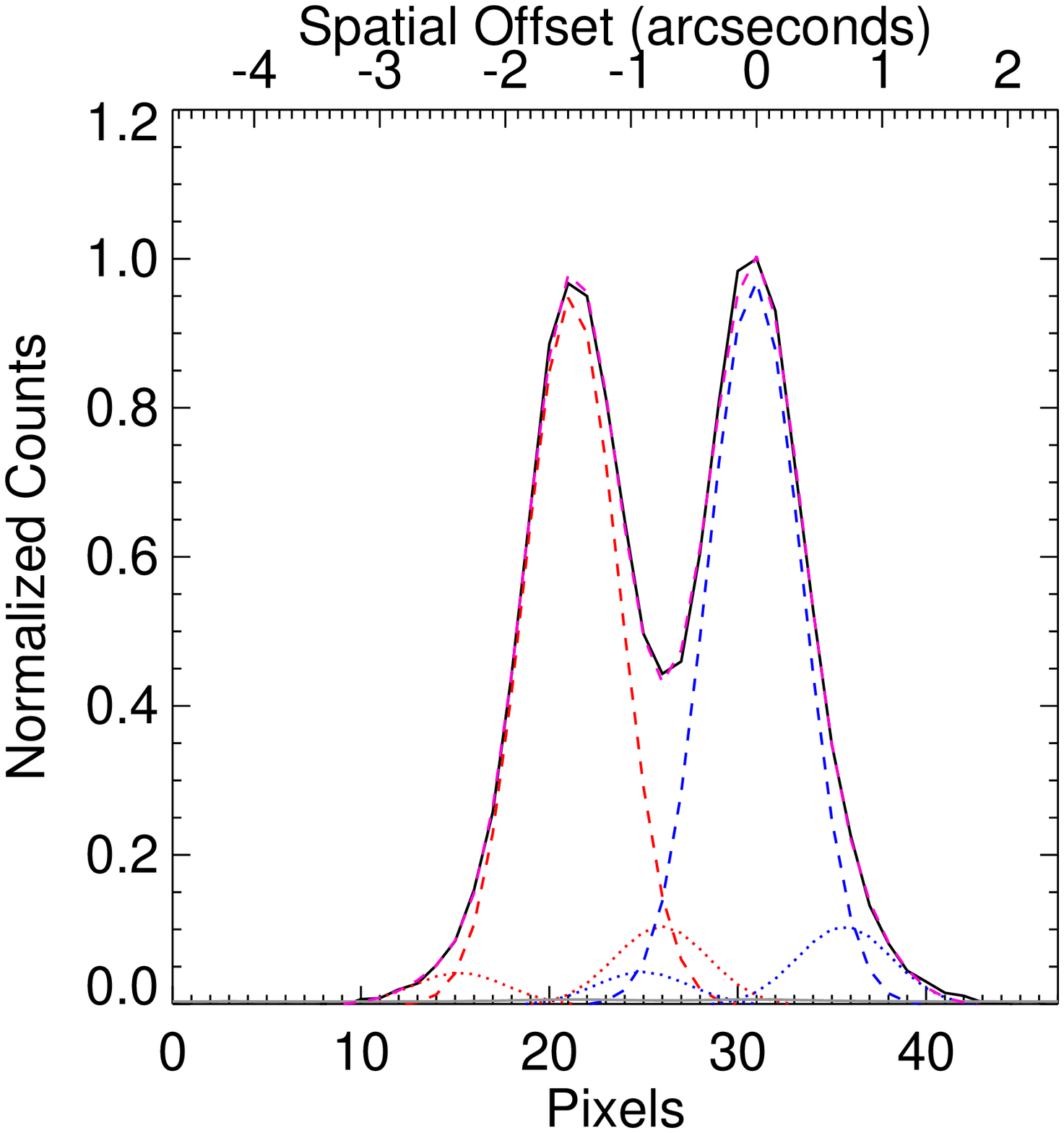}{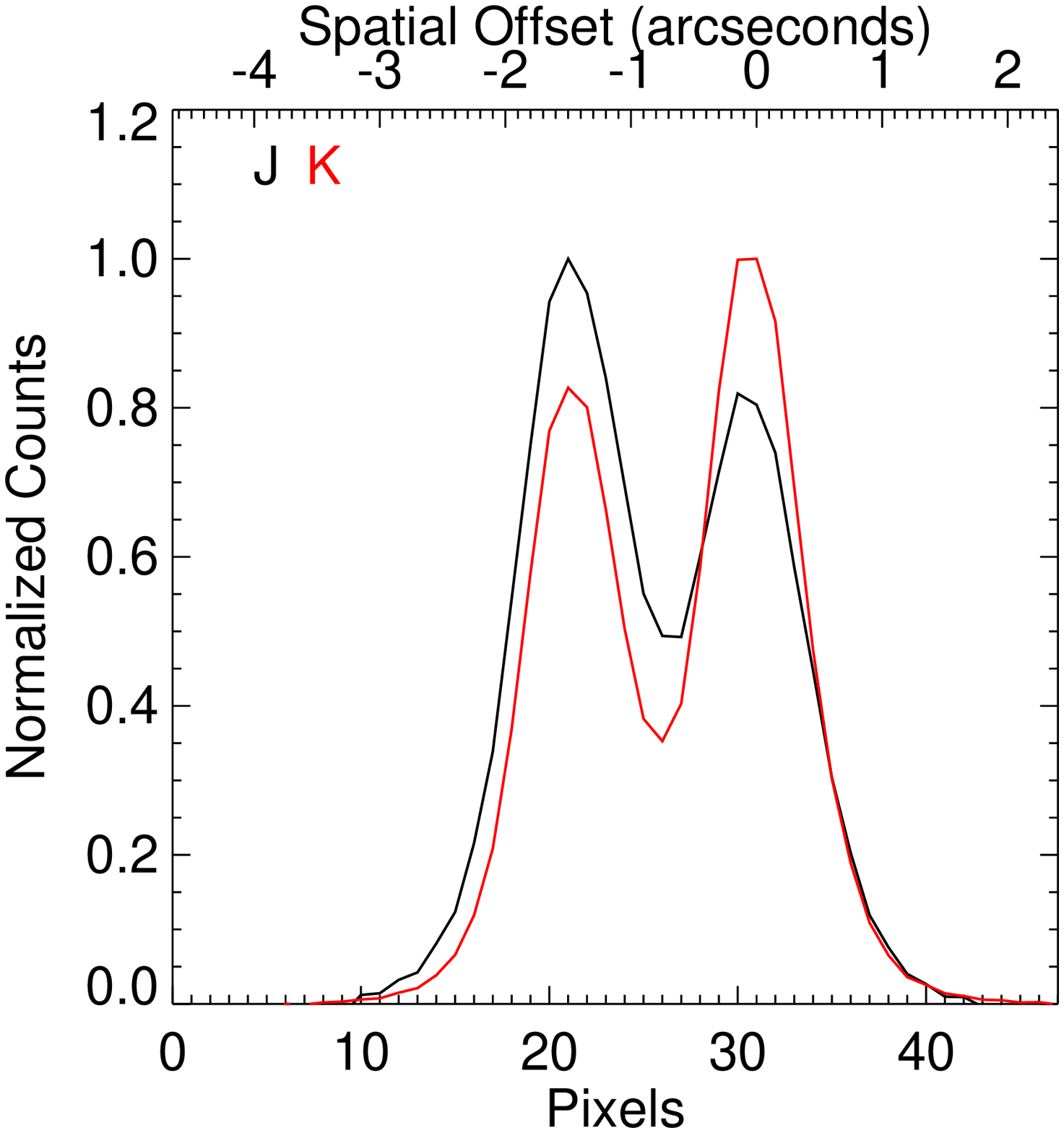}
\caption{(Left): Normalized spatial profile from the image in Figure~\ref{fig:model_image} integrated over wavelengths 1.0--1.3~$\micron$, 1.55--1.75~$\micron$ and 2.05--2.3~$\micron$ (black) compared to model profile (purple) with gaussian sub-components for primary (red) and secondary (blue) shown.  The profile uncertainties (grey) are $<$1\% and undetectable on this plot.
(Right): Comparison of normalized spatial profiles at $J$-band (black; 1.0--1.3~$\micron$) and $K$-band (red; 2.05--2.3~$\micron$), illustrating the flux reversal between the components.
}
\label{fig:model_profile}
\end{figure}

Figures~\ref{fig:model_image} and ~\ref{fig:model_profile} illustrate the quality of these fits, comparing the observed and modeled profiles as well as data and model images.
When the observed seeing was below 1$\farcs$2 (which encompassed 49 images in the period 6:03--6:46 UT), the profile fits converged exceptionally well, with fit residuals and corresponding spectral uncertainties typically 0.5-1\% in the brightest spectral regions. The modeling was also generally resistant to bad pixels, with the exception of a spectral detector crack visible in the 2.36--2.40~$\micron$ region in Figure~\ref{fig:model_image}.  

\begin{figure*}[t]
\center
\epsscale{1.0}
\plottwo{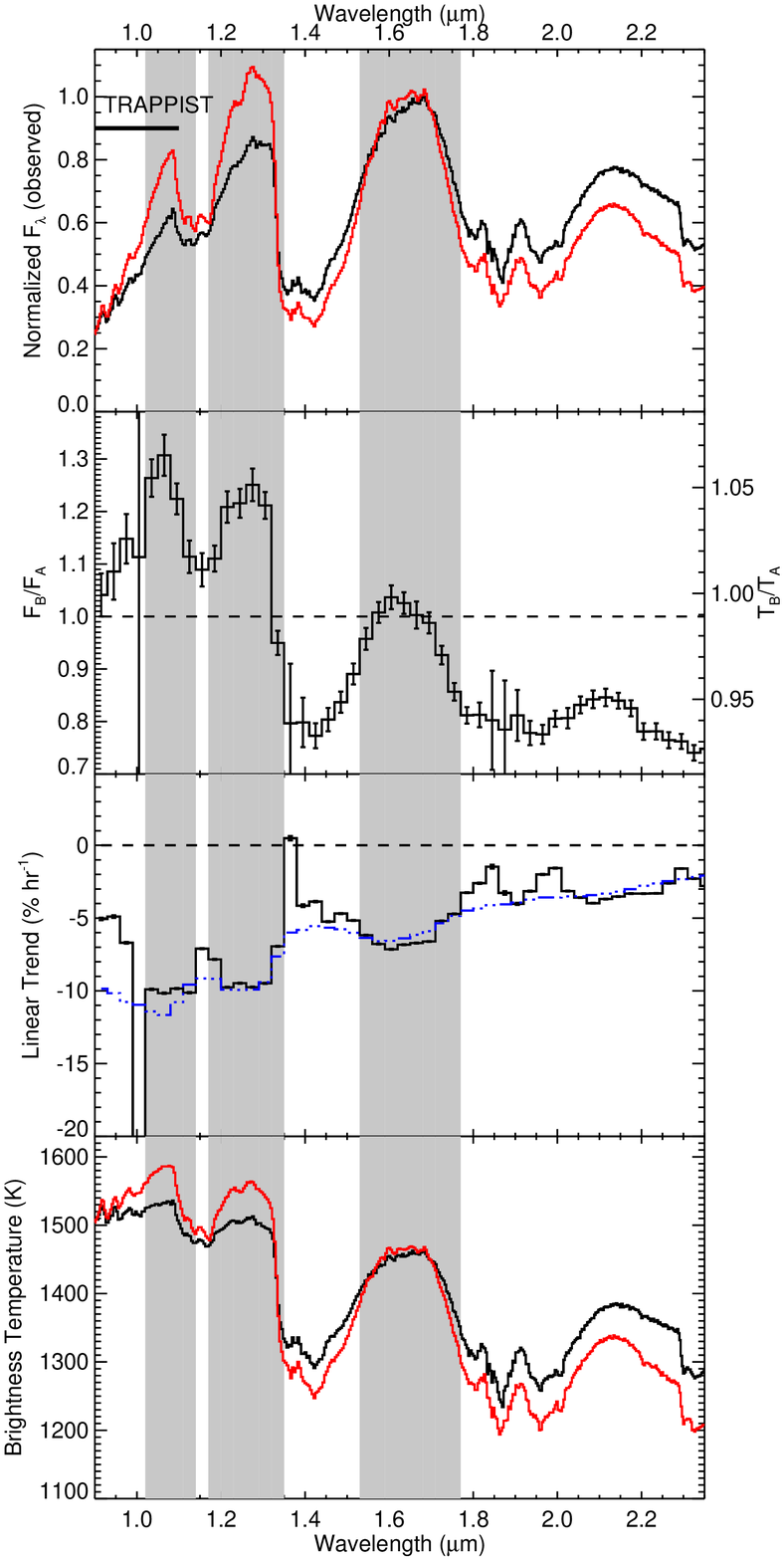}{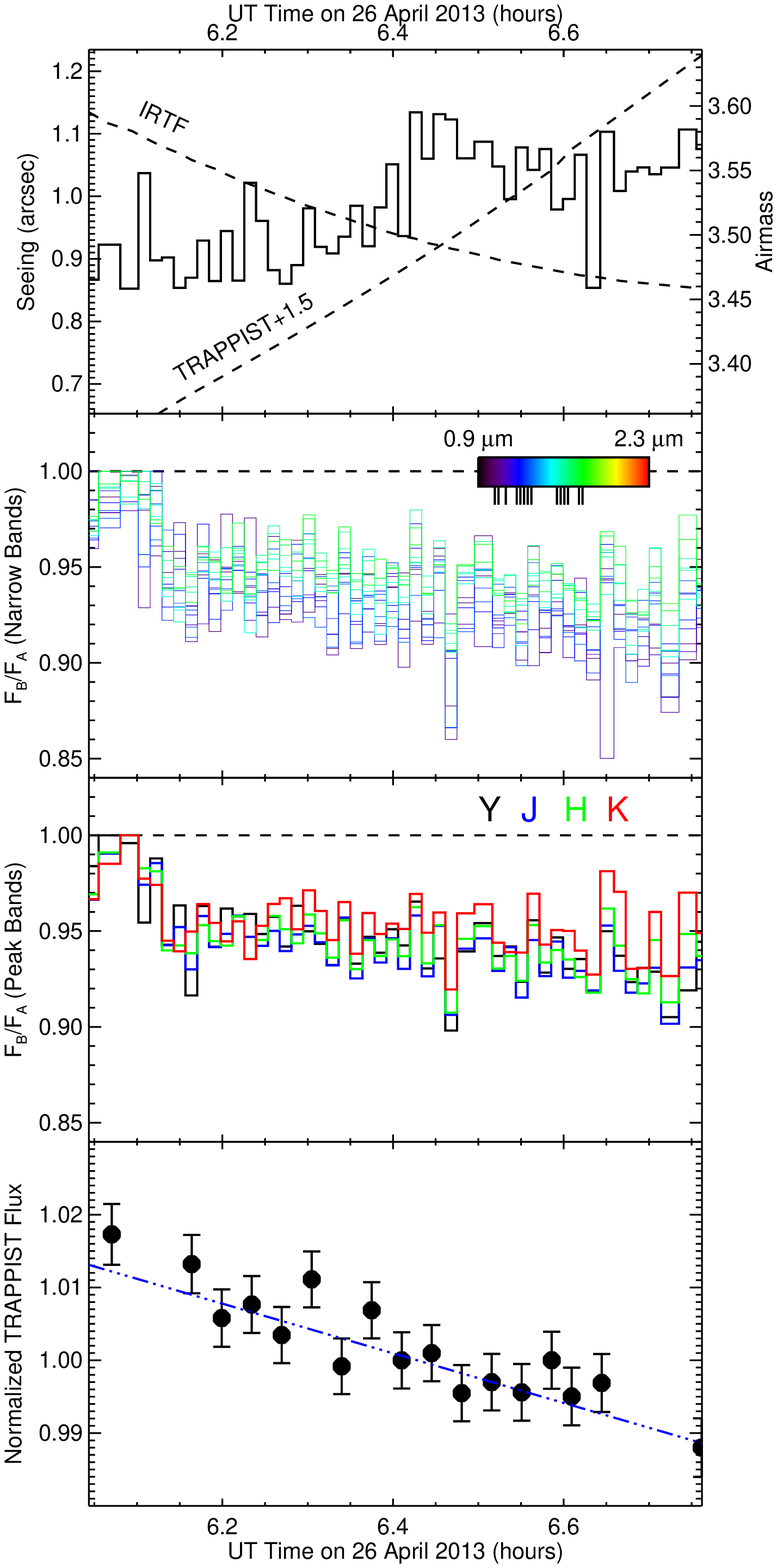}
\caption{Variability analysis from spectral (left) and temporal (right) perspectives.  
(Left Top): Observed fluxes for Luhman 16A (black) and B (red), averaged over the observing period and normalized by a common factor. { The TRAPPIST passband is indicated.}
(Left 2nd): Relative fluxes ($F_B/F_A$) averaged over the observing period, with error bars indicating the range of measured values. 
(Left 3rd): Inferred linear variations of relative fluxes as a function of wavelength during the observing period. The blue line represents the best empirical model fit. 
(Left Bottom): Brightness temperature spectra for Luhman 16A (black) and B (red).  In all panels the shaded regions indicate spectral regions with statistically significant variability based on the F-test.
(Right Top): Seeing for IRTF observations (black solid line, left axis) and airmass for IRTF and TRAPPIST observations (black dashed line, right axis) during the monitoring period. 
(Right 2nd): Time variation of $F_B/F_A$ in narrow spectral bands (0.03~$\micron$) with statistically significant variability.
(Right 3rd): Same as above but for broader band (0.1~$\micron$) $YJHK$ peak fluxes.  
(Right Bottom): TRAPPIST light curve over the same period, with the best-fit empirical model shown in blue.
}
\label{fig:variability}
\end{figure*}

Visually apparent even in the raw data, our observations confirm the flux reversal between {\namesh}A and B reported by \citet{2013ApJ...772..129B}.  As shown in Figure~\ref{fig:model_profile}, the T dwarf component is on average 20\% brighter at $Y$ and $J$ and 20\% fainter at $K$ compared to the L dwarf component.  Figure~\ref{fig:variability} displays these differences across the full spectral range, showing that {\namesh}B is the brighter component from 0.95-1.33~$\micron$, and marginally brighter even from 1.55--1.65~$\micron$.  Again, these are the spectral regions that are most influenced by condensate grain scattering and absorption in the L dwarfs, indicating that cloud opacity plays a primary role in the observed flux reversal.

\subsection{Relative Spectral Light Curves}

The observed fluxes are affected by three systematic effects: (1) slit losses due to the finite slit width used, which will vary with seeing and telescope tracking; (2) changes in atmospheric transmission due to the changing airmass over the observation; and (3) differential color refraction (DCR) induced by observing the pair aligned along their binary axis rather than the parallactic angle.  The last two factors are particularly problematic for this observation given the large airmass at which {\namesh}AB was observed.  Rather than devise a model to compensate for these effects, potentially introducing new systematic errors, we focused our analysis on relative flux variations over narrow spectral bands. This choice mitigates slit loss and transmission variations which affect both sources equally, and we assume that DCR does not induce significant color variation over a sufficiently narrow wavelength range.\footnote{Because these observations were taken at large zenith angle (18$\degr$ off the horizon), we cannot implicitly assume DCR effects are negligible in the near-infrared; see \citet{1996PASP..108.1051S}.}  

We quantified variability by measuring the relative observed fluxes of the two components, $R(\lambda,t) = F_B(\lambda,t)/F_A(\lambda,t)$, in 0.03~$\micron$ bands ($\approx$2--3 resolution elements) between 0.9--2.3~$\micron$
(Figure~\ref{fig:variability}).  We used the $\chi^2$ statistic to assess the presence of variability for each spectral band over the observing period:
\begin{equation}
\chi^2(\lambda) = \sum_{i=1}^{N_{obs}}\frac{(R(\lambda,t_i)-R_{\rm model}(\lambda,t_i))^2}{\sigma^2(\lambda,t_i)}.
\end{equation}
Here, $N_{obs}$ = 49, $\sigma(t)$ is the uncertainty in relative flux at time $t_i$ (typically 1--3\%) and $R_{\rm model}(\lambda,t)$ is the modeled value.  We considered 
{ the two simplest} models of constant flux ($R_{\rm model}(\lambda,t)$ = $R_0(\lambda)$) and linear variation with time ($R_{\rm model}$ = $R_0(\lambda)$ + $\alpha(\lambda)$t), and found that the latter was a statistically significant better fit to the timeseries data in the pseudocontinuum regions based on the 
F-test statistic (confidence of 95\% or greater).  Figure~\ref{fig:variability} displays the linear slopes in percentage change per hour as a function of wavelength.  Nearly all of the regions for which time variability is statistically significant are those in which {\namesh}B is the brighter component and, assuming identical radii, has a higher brightness temperature\footnote{To compute T$_{br}$ values, we first determined the scaling term for {\namesh}A that converts its apparent spectral flux to surface flux based on its measured absolute $J$-band magnitude of 15.00$\pm$0.04 \citep{2013ApJ...772..129B,2013arXiv1312.1303B} and a radius of { 0.86$\pm$0.06}~R$_{Jupiter}$ { using the evolutionary models of \citet{2001RvMP...73..719B} and assuming {\teff} $\approx$ 1300--1500~K and an age of 1-5~Gyr.} We applied the same scaling to {\namesh}B.  Brightness temperatures were assigned by determining the Planck blackbody that provides an equivalent flux density. The absolute brightness temperature values have uncertainties of 4\% based on uncertainties in the absolute magnitudes and radii of the sources, but relative temperature differences between the two components are certain to better than 0.5\%.} ($T_{br}$). This result is consistent with the results of \citet{2013A&A...555L...5G} and \citet{2013ApJ...778L..10B}, { who find} that {\namesh}B dominates the observed variability of the system. 
The magnitude of $\alpha(\lambda)$ decreases with increasing wavelength, from 10\%~hr$^{-1}$ at 1~$\micron$ to 4\%~hr$^{-1}$ at 2.1~$\micron$. This is qualitatively similar to broad-band photometric variations of SIPS~J0136+0933 \citep{2009ApJ...701.1534A} and 2MASS~J2139+0220 \citep{2012ApJ...750..105R}, which are observed to be greater at $J$ than $K$. On the other hand, the sense of this variation is also consistent with the declining brightness of {\namesh}B relative to the seemingly invariable   {\namesh}A \citep{2013ApJ...778L..10B}.  
{ We also note small declines in  $\alpha(\lambda)$ in regions of strong {\wat} absorption (1.35--1.45~$\micron$, 1.8--2.0~$\micron$), although signals-to-noise in these regions are much lower.}
 
We show the time series of the relative fluxes in significantly variable bands and in broader spectral regions (0.1~$\micron$) encompassing the 
$Y$ (1.12--1.22~$\micron$),
$J$ (1.25--1.35~$\micron$),
$H$ (1.6--1.7~$\micron$), and
$K$ (2.1--2.2~$\micron$) flux peaks in Figure~\ref{fig:variability}.
Remarkably, all of these regions show a common morphology: a fast ($\sim$3~min) dimming of order 5\% at UT 6:08, followed by a much slower decline for the remainder of the observing period.  The fast dimming does not appear to be related to sudden changes in seeing or airmass; indeed, a step-up in seeing at 6:24 UT does not coincide with any feature in the light curves. We verified that the decline beyond 6:08 remained significant for three regions in the $J$-band (1.095~$\micron$, 1.125~$\micron$ and 1.215~$\micron$) where {\namesh}B is brightest, with a linear declining trend of 5--6\%~hr$^{-1}$.   
Examining the broad-band spectral peak relative fluxes, we again see a wavelength dependence during the slow decline, with $F_B/F_A$ changing the most at $Y$ and $J$ bands and the least at $K$.  
Overall, it appears that the relative fluxes of these two components underwent a sharp then gradual decrease over the observing period, amounting to a $\approx$7.5\% ($\approx$5\%) decline in brightness at $J$ ($K$) over 45~min.
Note that our limited time coverage prevents assessment of the $\sim$100$\degr$ phase difference between $J$ and $K$ variations reported by \citet{2013ApJ...778L..10B}.

The decline in relative spectral fluxes aligns well with a decline in combined red optical light as measured by TRAPPIST (hatched region in Figure~\ref{fig:trappist}). The spectral monitoring period coincided with a 2.5\% decrease in total brightness, or a 4\%~hr$^{-1}$ linear trend with time, shallower than our near-infrared spectral band measurements.  
Since a decline in relative flux must be caused by a dimming secondary and/or brightening primary, and a decline in total flux by a dimming secondary and/or dimming primary, we logically conclude from both of these datasets that {\namesh}B is the variable component, in agreement with \citet{2013A&A...555L...5G} and \citet{2013ApJ...778L..10B}.

\subsection{An Empirical Model of the Observed Spectral Variability}

If {\namesh}B is the primary variable in this system, the wavelength dependence of the observed spectral variations, particularly in the pseudocontinuum regions where they are significant, arises from three possible effects. First, achromatic changes in the pseudocontinuum caused by pulsation and/or achromatic opacity variations, that manifest as a wavelength-dependent variation due to the changing relative fluxes of the two components across the near-infrared; second, chromatic variations arising from changes in intrinsically wavelength-dependent opacities; and third, a combination of both. To assess the underlying nature of {\namesh}B's variability, we used a simple empirical model to replicate both SpeX and TRAPPIST observations during the monitoring period.
Assuming {\namesh}A was invariable in the near-infrared during the time of our observations ($<$0.3\% variability was reported by \citealt{2013ApJ...778L..10B} in 4~hr of observation), and that the variability of {\namesh}B is linear (or nearly so) in time and/or wavelength over the period observed, { the simplest model for the spectrum of {\namesh}B taking into account these effects} is:
\begin{equation} 
F_B(\lambda,t) = F_{B,0}(\lambda) \times [a_0 + a_1(t-t_0) + a_2(\lambda-\lambda_0) + a_3(t-t_0)(\lambda-\lambda_0)].
\label{eqn:varmodel}
\end{equation}
Here, $F_{B,0}(\lambda)$ is the median spectrum of {\namesh}B over the monitoring period, $t_0$ is  the start of the period, $\lambda_0$ = 1.77~$\micron$ is the median wavelength of the spectrum, and the parameters $a_0$, $a_1$, $a_2$ and $a_3$ are linear coefficients taking into account relative scaling, achromatic time variation, chromatic scaling and chromatic time variation, respectively.   Assuming $F_A(\lambda,t) = F_{A,0}(\lambda)$, we used this function to calculate the linear rate of change of the relative spectra ($\alpha(\lambda)$), as well as the combined light of the system integrated over a constant 0.75--1.1~$\micron$ passband to simulate the TRAPPIST data. 

Applying the Nelder-Mead algorithm with a $\chi^2$ evaluation, we determined the best parameters for Equation~\ref{eqn:varmodel} fitting only for $\alpha(\lambda)$. We also performed fits in which one or more parameters were forced to be zero to assess their significance.  The best-fit model, shown in Figure~\ref{fig:variability}, required all terms except $a_2$, with $a_0$ = 1.39, $a_1$ = $-$0.0549~hr$^{-1}$ and $a_3$ = 0.0468~hr$^{-1}$~$\micron^{-1}$.  Fitting with $a_2$ gave similar values for the other parameters but made the overall fit slightly worse.  Excluding either the achromatic ($a_1$) or chromatic ($a_3$) variation terms, or both, produced significantly worse fits which could be excluded at  $>$95\% confidence based on the F-test statistic.

We may therefore conclude that both achromatic and chromatic pseudocontinuum variations were present in {\namesh}B during the monitoring period, variations that are consistent with changes in the cloud covering fraction \citep{2001ApJ...556..872A}. The positive value of $a_3$ is particularly relevant here, as it indicates that the variable opacity source plays a greater role at shorter wavelengths where gas opacity is minimal, as expected if { that source is condensate grain opacity}.
The predicted TRAPPIST light curve for the model { constrained by the SpeX spectra} also agrees well with that data (Figure~\ref{fig:variability}). { In particular, the model produces a} smaller amplitude of { optical} variation due to the { reduced} contribution of {\namesh}B to the combined light of the system { at these wavelengths}.

\section{Discussion}

\subsection{{ Interpreting} the Nature of {\namesh}B's Spectral Variability}

Spectral trends in variability have been examined in several L and T dwarfs to date, through pure spectroscopy (e.g., \citealt{2008MNRAS.384.1145B,2008A&A...487..277G,2013ApJ...768..121A}) and simultaneous or near-simultaneous broad-band imaging (e.g., \citealt{2004MNRAS.354..466K,2009ApJ...701.1534A,2012ApJ...760L..31B,2012ApJ...750..105R,2013AJ....145...71K,2013ApJ...767..173H,2013ApJ...778L..10B}).  The most significant variables up until now, SIMP~J0136+0933 and 2MASS~J2139$-$0220, both exhibit color trends in { near-infrared} photometric variability, with larger amplitude changes at $J$ as compared to $K$, again consistent with variable condensate cloud opacity.  However, spectroscopic variability measurements of these same two sources over 1.0--1.7~$\micron$ by \citet{2013ApJ...768..121A} indicate that achromatic or near-achromatic variations dominate the psuedocontinuum.  These authors propose { a two-layer cloud model with a thick shallow cloud and thin deep cloud as a means of reproducing} both achromatic psuedocontinuum and chromatic broadband variability.  
{ Matched to atmosphere models, this framework can replicate observed trends in the colors and spectral shapes of SIMP~J0136+0933 and 2MASS~J2139$-$0220 over 1-3 rotation periods, although detailed fits to the data remain poor (see also \citealt{2012ApJ...750..105R}).} 

For {\namesh}B, we also find that both achromatic and chromatic variations must be present in the psuedocontinuum to properly model the observations.  Achromatic variation yields a decline in the overall flux, amounting to roughly 0.03~mag in broad-band $J$ over the observing period. The concurrent chromatic variation simultaneously reddens the spectrum of this source by $\Delta(J-K)$ = 0.02~mag, resulting in a relative flux variation amplitude of $\Delta{F_{K_s}}/\Delta{F_J}$ = 0.41$\pm$0.18, similar to values reported for SIMP~J0136+0933 \citep{2009ApJ...701.1534A} and 2MASS~J2139$-$0220 \citep{2012ApJ...750..105R}. Combined, the achromatic and chromatic terms nearly cancel in the $K$-band, a region that is gas opacity dominated (H$_2$O, CH$_4$ and H$_2$). Hence, { our linear spectral model} is functionally consistent with condensate clouds being the primary driver of variability in {\namesh}B.

\subsection{ A Brightness Temperature Spot Model for Luhman~16B}

{ Given the known shortcomings in reproducing the near-infrared spectra of L/T transition brown dwarfs (e.g., \citealt{2008ApJ...682.1256L,2008ApJ...678.1372C,2009ApJ...702..154S}), we forgo detailed modeling of the spectra in lieu of a simply brightness temperature variation model, focusing at 1.25~$\micron$ where gas opacity is a minimum and cloud structure variations are expected to have the greatest influence \citep{2001ApJ...556..872A}.  
The simplest model for replicating the surface flux $\langle{F}\rangle$ of a patchy brown dwarf is two sets of regions with differing brightness temperatures covering the surface:}
\begin{equation}
\langle{F}\rangle \propto \langle{T_{br}}^4\rangle \equiv {A}T_{cold}^4 + (1-A)T_{hot}^4.
\label{eqn:tsurf}
\end{equation}
{ Here, $\langle{T_{br}}^4\rangle$ is the disk-averaged brightness temperature,} 
$T_{cold}$ and $T_{hot}$ are the brightness temperatures of cold and hot { regions}, respectively, $A \equiv F_{cold}/\langle{F}\rangle \leq 1$ is the areal covering fraction of the cold { regions}, and we ignore limb darkening.  
Our interpretation of this model is that the cold { regions} correspond to the { highest cloud layer in the brown dwarf atmosphere}, while the hot { regions} correspond to gaps in these clouds  that probe to some as-yet undetermined deeper layer with brightness temperature $T_{hot}$; 
{ a cartoon perspective of this is shown in Figure 6 of \citet{2013ApJ...768..121A}. We note that this is not the only interpretation of a two-spot model, which could also arise from magnetic interaction at the photosphere (i.e., starspots) or updrafts of warm air pockets driven by convective flows. Nevertheless, we will occasionally refer to the cold { region} as ``clouds'' and hot { regions} as ``holes'' in the following discussion.}

\begin{figure}[h]
\center
\epsscale{1.0}
\plotone{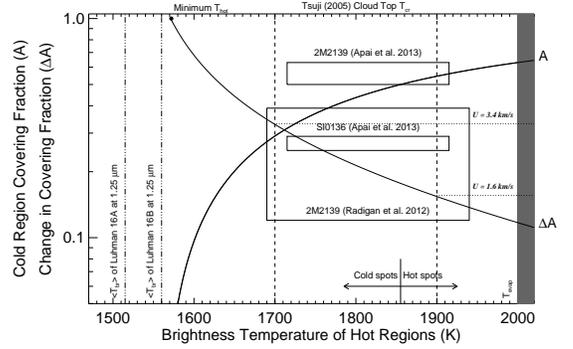}
\caption{{ Cold spot} covering fraction $A$ and change in covering fraction $\Delta{A}$ (solid lines) on {\namesh}B { over a full cycle}, as a function of the hot spot brightness temperature at 1.25~$\micron$. Values for $A$ are derived from Eqn.~\ref{eqn:tsurf} and assume $\langle{T_{br}}\rangle_{L16B}$ = { 1560~K} and $T_{cold}$ = ${\langle}T_{br}\rangle_{L16A}$ = { 1510~K}, based on Figure~\ref{fig:variability} (vertical triple-dot dash lines).  Values for $\Delta{A}$ are derived from  Eqns.~\ref{eqn:da1} and~\ref{eqn:da2} and assume $\Delta{F}/F$ = 13.5\%.  The point where $\Delta{A}$ = 100\% indicates the minimum $T_{hot}$; { the temperature at which $A = 0.5$ is also labeled, separating cold surfaces with hot spots from a hot surface with cold spots.} Also labeled are the range of $T_{cr}$ values from \citet{2005ApJ...621.1033T} and \citet{2012ApJ...760..151S} that estimate the height of cloud tops, and the temperature at which all condensates are assumed to be evaporated ($T_{evap}$).  Estimates for $A$ and $T_{hot}$ for SIMP~J0136+0933 and 2MASS~J2139$-$0220 from \citet{2012ApJ...750..105R} and \citet{2013ApJ...768..121A} are indicated, assuming the same cloud top temperature.  Finally, we label estimates of jet size scales for wind velocities of { $U$ = 1.6~{\kms} and 3.4~{\kms} based on a Rhines length scale (Eqn.~\ref{eqn:rhines}); these intersect the $\Delta{A}$ curve at $T_{hot}$ = 1700~K and 1900~K}.
}
\label{fig:patchy}
\end{figure}

For {\namesh}B,  ${\langle}T_{br}\rangle$ = { 1560~K} at the 1.25~$\micron$ $J$-band peak continuum (Figure~\ref{fig:variability}; see also \citealt{faherty14}).  If we take the brightness temperature of {\namesh}A at this wavelength, { 1510}~K, as an estimate for $T_{cold}$ for both sources,\footnote{The assumption can be justified in part by the nearly identical brightness temperatures of the two sources in the 1.15~$\micron$ region, where {\wat} and {\meth} opacity play a larger role than clouds.} then we can jointly constrain $A$ and $T_{hot}$, as illustrated in Figure~\ref{fig:patchy}.  { Coverage of cold regions} is essentially negligible for $T_{hot} <$ 1570~K, then climbs to over 50\% at $T_{hot} \approx$ 1860~K.  At hotter temperatures, { our model predicts that the atmosphere of {\namesh}B would be overall similar to that of {\namesh}A with occasional hot spots, which we assume to be less than the evaporation temperature of mineral condensate species} ($T_{evap} \approx$ 2000~K; \citealt{1999ApJ...519..793L}).  
We note that { equal hot-cold spot coverage for {\namesh}B  occurs in} the 1700--1900~K range that \citet{2005ApJ...621.1033T} estimate as the effective top of a brown dwarf cloud layer ($T_{cr}$; see \citealt{2012ApJ...760..151S}).  
\citet{2012ApJ...750..105R} and \citet{2013ApJ...768..121A} also provide estimates for $A$ and $\Delta{T}_{hc} = T_{hot}-T_{cold}$ for SIMP~J0136+0933 and 2MASS~J2139$-$0220 based on their own { two-spot modeling}. While temperature values reported in these studies are based on model effective temperatures, if we assume that the brightness temperature offsets are the same as their preferred $\Delta{T}_{hc} \approx$ 300~K, this would also place the hot { regions of {\namesh}B in the same temperature range as} the cloud tops of the \citet{2005ApJ...621.1033T} models. 
{ Thus, if our spot model is interpreted as probing different layers of {\namesh}B's atmosphere, the best estimates of the temperature differential is in line with the conjecture of \citet{2013ApJ...768..121A} that gaps in the highest cloud deck still probe regions influenced by condensate opacity.  
However, we stress that our data cannot independently determine $A$ or $T_{hot}$, and other interpretations of these temperature differences are conceivable.}

The fractional peak-to-peak variation in observed flux that occurs as { hot and cold regions} rotate in and out of view is
\begin{equation}
\frac{\Delta{F}}{\langle{F}\rangle} = \frac{\Delta{A}}{A-\epsilon}
\label{eqn:da1}
\end{equation}
where $\langle{F}\rangle$ is the average flux, $\Delta{F}$ the change in total flux, $\Delta{A}$ the change in cloud coverage (increasing $A$ decreases the total flux), and $\epsilon \equiv F_{hot}/(F_{hot}-F_{cold}) = T_{hot}^4/(T_{hot}^4-T_{cold}^4) \geq 1$; see also \citet{2012ApJ...750..105R}.\footnote{Our expression differs slightly from \citet{2012ApJ...750..105R} because we assign $A$ to be the cloud-covering fraction, whereas they define the equivalent parameter $a$ as the cloud-cleared fraction.}
In terms of brightness temperatures:
\begin{equation}
\Delta{T_{br}}  =  \frac{\Delta{F}}{4\langle{F}\rangle}\langle{T_{br}}\rangle = \frac{\Delta{A}}{4(A-\epsilon)}\langle{T_{br}}\rangle.
\label{eqn:da2}
\end{equation}
During our observations, we observed a 7.5\% variation in the $J$-band peak continuum that was coincident with a 2.5\% variation in TRAPPIST red-optical photometry. We therefore assume that the full 4.5\% peak-to-peak variation in TRAPPIST photometry around our spectral observations (Figure~\ref{fig:trappist}) corresponds to a 13.5\% variation at $J$, or a peak-to-peak temperature fluctuation of $\Delta{T_{br}} \approx$ 50~K\footnote{The maximum variation observed by \citet{2013A&A...555L...5G} over weeks of monitoring is 10\%, which would correspond to 30\% variations in $J$, exceeding those observed in 2MASS~J2139$-$0220. However, we restrict our analysis here to the period around the spectral observations { since the spectral response of larger fluctuations may differ}.} following Eqn~\ref{eqn:da2}.  This temperature offset is notably similar to the temperature difference between {\namesh}A and B at these wavelengths (Figure~\ref{fig:variability}). Using the { relationship between} $A$ and $T_{hot}$ above, we computed $\Delta{A}$ as a function of $T_{hot}$, also shown in Figure~\ref{fig:patchy}.  Not surprisingly, the { areal} variation required to reproduce the observed brightness variations declines with higher $T_{hot}$; i.e., with greater contrast between cold and hot { regions}.  
An important reference point is the temperature at which { areal} variations become smaller than the total { cold region} coverage, which occurs for $T_{hot} >$ 1710~K and $A >$ 30\%. { The corresponding $\Delta{T}_{hc}$ = 150~K is on the low end of estimates} for  SIMP~J0136+0933 and 2MASS~J2139$-$0220, and just above the minimum $T_{cr}$ from \citet{2005ApJ...621.1033T}. { For the range 1700~K $< T_{hot} <$ 1900~K, which we again take as a reasonable estimates of the hot spot temperature, the inferred cold covering fraction is roughly 30--55\%, intermediate between similar values inferred for SIMP~J0136+0933 (25-30\%) and 2MASS~J2139$-$0220 (50-65\%) by \citet{2013ApJ...768..121A}.

\subsection{ Interpretation: Rhines Length Scale and Advective Time Scale}

For 1700~K $< T_{hot} <$ 1900~K, { cold spot coverage} must vary by 15--30\% over a single period to replicate the observed variability amplitude, { implying a $\sim$ 30-100\% variation between hemispheres if the spot patterns are static.
Organized jet features in the atmospheres of the giant Solar planets generally scale in size with the 
Rhines length \citep{1970GApFD...1..273R,2008ASPC..398..419S}, $L_{Rh} \sim \left({U}/{2\Omega{R}\cos{\phi}}\right)^{1/2}$, where $U$ the characteristic wind speed, $R$ is the radius, $\Omega= 2\pi/P$, $P$ is the rotation period and $\phi$ is the latitude of the feature.  If we assume that the same scaling occurs for features in brown dwarf atmospheres (e.g., \citealt{2013ApJ...768..121A,2013ApJ...776...85S}), then their maximum fractional size scale is:
\begin{equation}
\alpha_{Rh} \sim \left(\frac{L_{Rh}}{R}\right)^2 \approx  2\%\left(\frac{U}{\rm km/s}\right)\left(\frac{P}{\rm hr}\right)\left(\frac{R_{Jup}}{R}\right).
\label{eqn:rhines}
\end{equation}
where we have assumed mid-latitude features. If we now relate this maximum scale to the areal spot variation inferred here ($\alpha_{Rh} \sim \Delta{A}$), the known rotational period and assumed radius of {\namesh}B implies characteristic wind speeds of 1.6~{\kms} $< U <$ 3.4~{\kms} for 1700~K $< T_{hot} <$ 1900~K (Figure~\ref{fig:patchy}). These speeds are somewhat higher than the range favored by the circulation models of \citet{2013ApJ...776...85S}, assuming winds are driven by inefficient conversion of convective heat (10--300~{\ms})}.   
However, the speeds do give advection timescales, $\tau_{adv} \sim R/U \sim (2-5)\times10^4~s \sim 1-3$ rotation periods, that are consistent with the timescale of lightcurve evolution observed in {\namesh}B  \citep{2013A&A...555L...5G}. 

The convergence between the inferred variation and Rhines length scales, and the advective and evolutionary time scales, suggest that our gross estimates for $T_{cold}$, $T_{hot}$, $A$ and $\Delta{A}$ are not too far off the mark.  However, we have made a number of major assumptions that require confirmation through more detailed spectroscopic monitoring and modeling, in particular to ascertain whether the spot regions have spectral characteristics (features and line profile shapes) consistent with the inferred brightness temperatures.  Nonetheless, our basic model of a cold cloud deck disrupted by warm dynamic features shows promising agreement with planetary analogs and current brown dwarf circulation models.}

\subsection{Trends in L/T Transition Variability}

\begin{deluxetable*}{lcccccl}
 \tablecaption{Comparison of Highly Variable L/T Transition Dwarfs. \label{tab:comp}}
 \tabletypesize{\small}
 \tablewidth{0pt}
 \tablehead{
 \colhead{Source} &
 \colhead{SpT} &
 \colhead{$P_{rot}$} &
 \colhead{$\Delta{F}/F$} &
 \colhead{$A$} &
 \colhead{$\Delta{K}/\Delta{J}$} &
 \colhead{Ref} \\
  & & 
 \colhead{(hr)} &
 \colhead{at 1.25~$\micron$} \\
}
 \startdata
SIMP~J0136+0933	& T2.5 & 2.3895$\pm$0.0005 & 5.5\% & 25-30\% & 0.48$\pm$0.06 & 1,2 \\
{\namesh}B  & T0.5 & 4.87$\pm$0.01 & 13.5\%\tablenotemark{a} &  30-55\%\tablenotemark{b} & 0.41$\pm$0.18 &  3,4 \\
2MASS~J2139$-$0220 & T1.5 & 7.721$\pm$0.005 & 30\% & 50-65\% & 0.45--0.83 & 2,5 \\
\enddata
\tablenotetext{a}{Based on the maximum peak-to-peak TRAPPIST variability amplitude during the current observing period.}
\tablenotetext{b}{Assuming $T_{hot}-T_{cold}$ = 300$\pm$100~K; see \citet{2013ApJ...768..121A}.}
\tablerefs{
(1) \citet{2009ApJ...701.1534A};
(2) \citet{2013ApJ...768..121A};
(3) \citet{2013A&A...555L...5G};
(4) This paper;
(5) \citet{2012ApJ...750..105R}.}
\end{deluxetable*}

{\namesh}B joins SIMP~J0136+0933 and 2MASS~J2139$-$0220 as the three most variable L/T transition objects detected to date, so it is worth comparing the variability properties of these sources, summarized in Table~\ref{tab:comp}.   The variability period, $J$-band variability amplitude and inferred cloud covering fraction
of {\namesh}B are all intermediate between those of SIMP~J0136+0933 and 2MASS~J2139$-$0220, although epoch-to-epoch changes in these values are considerable.  As the Rhines scale scales linearly with the rotation period,\footnote{\citet{2013ApJ...768..121A} incorrectly state a spot scaling law of ${P}^{-2}$ in the text, but infer a spot scaling between 2MASS~J2139$-$0220 and SIMP~J0136+0933 that is consistent with $A \propto P$; the former is likely a typographical error.}, 
{ its interpretation as an estimate of surface feature size is consistent with {\namesh}B's intermediate period and intermediate variability amplitude, as a few large features are more likely to give rise to stronger disk-integrated variations than many small features \citep{2013ApJ...768..121A}.} 
There also appears to be a correlation between rotation period and cloud covering fraction, although temperature effects may play a role in this statistic. The source with the smallest cloud coverage, SIMP~J0136+0933, is also the latest-type and presumably coldest brown dwarf in the sample.  Finally, we find essentially no difference in color variability among these sources. As noted above, our estimate of $\Delta{F_{K_s}}/\Delta{F_{J}}$ for {\namesh}B is consistent with similar measures for SIMP~J0136+0933 and 2MASS~J2139$-$0220 { (although the latter can exhibit more extreme color terms; \citealt{2012ApJ...750..105R}), suggesting that the condensate clouds responsible for the variations in these sources are likely to} have similar opacities and physical properties (i.e., composition, grain size distribution, vertical structure, etc.). { However, confirmation of this agreement will again require more careful spectral modeling to accurately determine cloud properties.}

\section{Summary}

We have measured significant variability in the resolved, relative spectral fluxes of {\namesh}A and B using IRTF/SpeX.  Variations occur at all wavelengths, most significantly in the bands where {\namesh}B is the brighter source. We detect both a rapid decline of 5\% in about 3~min, and a subsequent slow decline in the remaining 45~min of observation, with rates ranging from $-$10\%~hr$^{-1}$ at 1.25~$\micron$ to $-$4\%~hr$^{-1}$ at 2.1~$\micron$. By comparing to concurrent combined-light photometry from TRAPPIST, we deduce that the observed variability originates from the T0.5 secondary, confirming the results of \citet{2013A&A...555L...5G} and \citet{2013ApJ...778L..10B}.  We are able to successfully reproduce both the SpeX and TRAPPIST lightcurves with an empirical model that assumes {\namesh}A is constant while {\namesh}B undergoes both achromatic and chromatic pseudocontinuum flux variations.  Qualitatively, this model { may be interpreted as arising from variations in cloud covering fraction in the photosphere of {\namesh}B as it rotates, although other physical models (starspots, gas upwelling) may also apply}. Using a simple { two-spot} model that assumes { cold regions are identical to the atmosphere of {\namesh}A}, we are able to deduce an average and variance in the { cold (or cloud)}  covering fraction of {\namesh}B as a function of the { temperature of hot (or hole) regions}. While the { hot region} temperature remains a free parameter, a range of 1700--1900~K is supported by the cloud models of \citet{2005ApJ...621.1033T} and the temperature contrasts inferred for SIMP~J0136+0933 and 2MASS~J2139$-$0220.
This range is also supported by the supposition that { surface features follow a Rhines scale, and predict wind velocities of 1--3~{\kms}. These are higher than early expectations from brown dwarf circulation modeling, but consistent with advection timescales that align with lightcurve variability over a few rotation periods.  Rhines scale-sized features also} explain the apparent trend between variability period and amplitude between SIMP~J0136+0933, {\namesh}B and 2MASS~J2139$-$0220.

The relative spectral fluxes of {\namesh}A and B, the presence of significant near-infrared variability on {\namesh}B, and the spectral nature of this variability all align with the model of cloud evolution through fragmentation as a driving mechanism for the L/T transition.  { However, other physical interpretations remain viable, and the influence of secondary parameters (surface gravity, metallicity, viewing perspective) are only starting to be explored \citep{2006ApJ...651.1166M,2010ApJ...710.1142B,2011ApJ...737...34M}.} Given its unique composition { and proximity to the Sun}, the {\namesh}AB system should continue to be monitored as a benchmark for cloud structure and evolution in cool brown dwarf { and exoplanet atmospheres}.

\acknowledgments

The authors thank Michael Cushing for assistance in the SpeX component extraction;
Dave Griep at IRTF for his assistance with the observations,
and helpful comments and contributions from Daniel Apai.
A.H.M.J. Triaud is a Swiss national science foundation fellow under grant PBGEP2-145594.
TRAPPIST is a project funded by the Belgian Fund for Scientific Research (Fonds National de la Recherche Scientifique, F.R.S- FNRS) under grant FRFC 2.5.594.09.F, with the participation of the Swiss National Science Foundation (SNF). M.\ Gillon and E.\ Jehin are FNRS Research Associates. This research has benefitted from the SpeX Prism Spectral Libraries, maintained by Adam Burgasser at \url{http://www.browndwarfs.org/spexprism}; and the M, L, T, and Y dwarf compendium housed at \url{http://DwarfArchives.org}.
We thank our anonymous referee for her/his comments.

Facilities: \facility{IRTF: SpeX}, \facility{TRAPPIST}

\end{document}